# Next Generation High Speed Computing Using Photonic Based Technology

[1]Umer Farooq, [2]M. Aqeel Iqbal, [1]Muhammad Ahsan and [1]Mashood Malik
[1]BCSE Students, DSE, Faculty of E & IT, FUIEMS, Rawalpindi, Pakistan
[2]DSE, Faculty of E & IT, FUIEMS, Rawalpindi, Pakistan
[2][DCE, College of E & ME, NUST, Pakistan]

*Abstract:* In the present era of technology computer has facilitated the human life up to a great extent. The speed of computation has raised to astonish level but the pace of development of other technologies which have core dependency over computers have raised relatively exponentially huge, though the computer speed of computation is very fast with respect to human abilities but still it has to be increased a lot more to meet the future requirements. We have pushed electrons to their maximum limit to a stage that nothing further could be expected from electrons. Alternately one can use photon to replace the relatively sluggish electrons. An alternate that posses all feature that an electron holds but only millions of time faster and with a far more reliability in one way or the other stretching the computers speed to a stage that no one would have ever even wonder. In this research paper the photonics implementations in computation industry have been presented along with its scope as an alternate to electron with comparative study of electron and photon under the computation perspective, generalized working of silicon based optical computers, the application of photons and their crucial role in the upcoming times.

## I. INTRODUCTION TO PHOTONIC TECHNOLOGY

Initialized as an enhanced amplifier no one could predict the extent to which the electronic transistor would revolutionize the technology. It yielded the micro electronics that paved the way for the luxuries of life we employ in our every day life with concurrent development and modifications periodically. Concurrently laser was discovered and later it revolutionized the music, printing and industrial application beyond the scope of practically possibilities domain of human mind. Right now we have achieved a mile stone in photonics , whose revolution is yet to be observed , photon for sure would be proving its worth by virtue of its inherit potentials and would be breaching all the current technological barriers currently electronic industry is bounded within. Photon is expected to replace electron as an alternate with ability to yield relatively exponentially huge data with far more reliability, speed and cost effectiveness. Photonic computing is considered to be turning point of technology same was as electronic was considered in twentieth century. It is wrong to claim photonics as emerging field because it has lot of products readily available in market. [1]

It was the primary target of the stakeholder of the computation industry to deliver the data fastly because of increasingly dependency of all aspects of human life activities based on computers. [1] Many approaches were adopted in order to achieve the targeted goals though it succeeded to a great extent but eventually as due to a certain electron potential barrier their was a limit set over it. Moreover any further enhancement was proving to be exponentially expensive though even if the change or so called increase in efficiency is very little. All these circumstances began to threaten the future of electronic industry and the quest to get an alternate for the drowning electronic industry initiated. [1][2]

Electrons may be very fast for present computations but what about the future when linguistic time would start with artificial intelligence, auto detection would not be a distinguish or an extraordinary activity rather would be a basic and often used activity. How will the present electronic based computation devices would be able to tackle such heavy processing a wonder of intelligent machines we dream of. Simply the answer is yes we will achieve our targets. [1][3][4]

In present conditions we can fit approximately up to 300 million transistors on a single chip. It is also estimated to be double in period of 18 to 24 months. Figure 1 shows the relation between the time period and the relatively increasing chip density. Now if we further try to enhance the IC performance we would have to increase the density of transistors on a chip, which are leading to serious compromises on the reliability of the devices. Though as for sake of overcoming these factors we have invented a variety of methods of increasing chip density ensuring no compromising on the working and reliability of device but all these measure are temporary and we would be facing them unless or until we find an alternate that will go beyond the barrier of electron. [1][3][5][6].

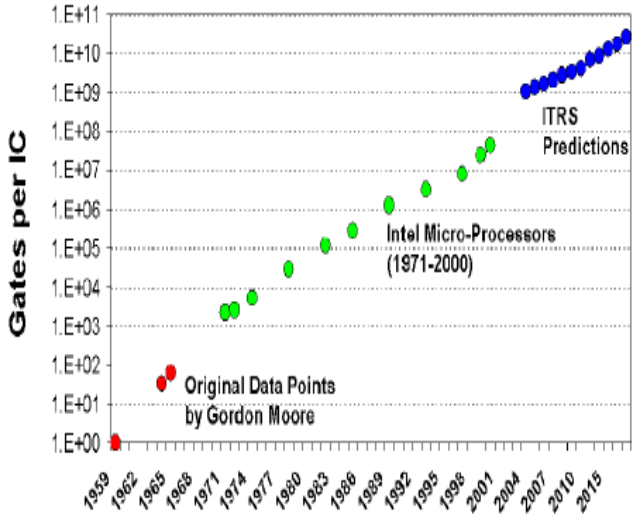

Figure 1. Moore's Law predictions for transistor density in chip. [6]

In the quest for a solution to the mentioned scenario and to achieve our dreamed future, the alternates for electrons were searched and eventually "Photon" was detected to be an alternate for the sluggish electron an less expensive but performance wise more valuable[1][2][4][5]

While the analysis of photonic instead of electrons, there's appear variety of criteria to analyze each other. As in photonic circuits, light would be used which becoming a sure is shot innovative for the 21st century technological development. As contrast to present electronic based computation, which is to identify presence referred "one" or absence referred "zero" one can perform computations at the rate of 2 BPT assuming in to the electron flow limit across system bus. On contrary or simply for ahead we can identify light (photon) in 16 distinguish states (categorically based on wavelength of light). Now if we use it we would get rate of power of computing to be 16 x BPT. Assuming the general limit of BPT speed to be 32 for electron based computing. It would be 232 or 4,294,967,296 BPT sound huge bits not even an ocean drop if we take the same BPT rate for photonic computer it would be 1632 or 340,282,366,920,938,463,463,374,607,431,770,000,000
BPT. Figure 2 reflect the optical communication between a photonic based CPU and its peripheral devices. That one could bet if one uses photons instead of electrons, to be straight forward. Photonics shall be replacing majority of the electronic industry and the main reason in contrast to photonic even the latest and most efficient electronic counterpart is far behind. [5]

All the present era wonders of photonics which are at their places state of the art are some achievement that no one ever thought would be achievable. But a brain clicks why its photon doing all this. These processes done by photons are for sure one 150-160 Gbps where it's just a joke to ask for electronic equivalent to perform so. [5]

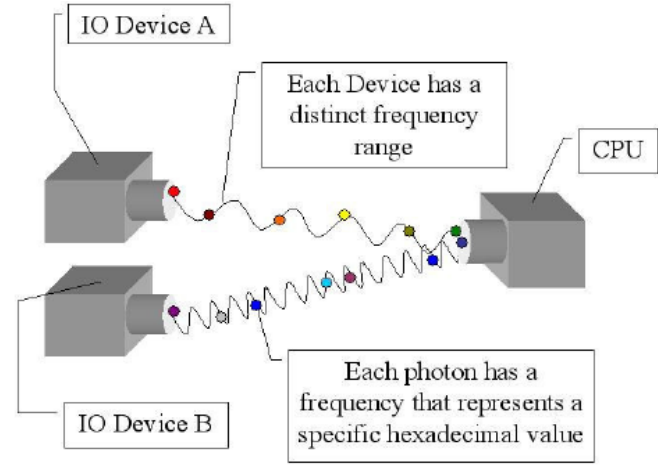

Figure 2. Communication of photonic components

All this is achieved by using a computer architecture entirely based on photonics. In doing so a light wave with a particular wavelength is given and the signal is then interpreted by the CPU. The photon sent makes the CPU that from where the light signal is from. In case of photon the light is divided into 16 band gapes based on difference in the wave length and each representing an entirely unique hexadecimal light.[3] Then the signal received by the CPU from the device would be photonic for example 14A3DC. The form of this signal would remain photonic until it reaches the processing chip via the optical medium generally optical bus. The signal would be interpreted and the desired task would be accomplished. All this would be achieved in a speed that one can't even dream of. [5][3]

II. SCOPE OF PHOTONIC TECHNOLOGY AND TOOLS

There are several achievements that we won't be making use of in our present while eliminating photonics contribution in them. Following are the main facts that reflect the photonic technology leap for ahead to an extent one even imagine:-[5] [8]

**A.** As photons are particles of light so the computation is directly dependent on speed of light which having a clear leap ahead to speed electrons. As the present photonic transistor can switch light equivalent to 1.5 fento seconds (one million of billionth hay a seconds) hence making full use of speed of light via processing the speed of light. [5][8]

**B.** Relatively to electron propagation which require field of energy to propagate from one position to the next so in circular path this is not at all matter of concern while using photons it would be following a straight path relative to electron circular path which gives photons an edge over electron. [3][8]

**C.** The more the technology progressing so is the pace of data transfer is required to be more quicker, however for electron there is a barrier after which the device performance relatively alters though the industry is trying its best and is overcoming it but how log will it last and the only way out is photon using which is no at all dependent on properties that put some physical barrier over it. [3][5][8]

**D.** The structure of electron and photons also alter for electrons its presence or absence there are two cases hence limited data. On contrary light having distinguished color wavelength has independent value. So it is like no of electrons hold together equivalent to a single beam of photon. [5][8]

**E.** Photonic transistors being component of electromagnetic spectrum. It's estimated that it could have 35 billion non-interacting separate channels referring if we use photon busses they shall be processing 35 billion pathways rather than present max of 64 or 128 KB buses. [3][5][8]. Figure 3 shows the classification of light on grounds of difference in wave lengths the portion indicted as communication is the optical domain suited for communication and presently being employed for optical fiber communication.

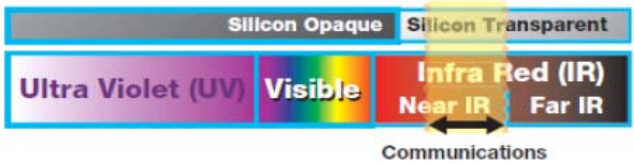

Figure 3. Fiber optics range studies for computations [4]

**F.** Non-practically colors can transmit 200 terabit of data. Now as out photonic transistor shall be operating on light, hence every device produce shall be processing huge amount with a light speed [3][5][8]

**G.** Holographic technique for photonics is far more efficient and liable than equivalent electron chips. [3][5][8]

**H.** Due to non-interacting within a property of light it could be shrinks far more that equivalent electronic chips. [3][8]

**I.** Physical components for photonic transistor are far more cheap with respect to electronic counter part e.g. glass, plastic, aluminum e.t.c with respect to pure semi conductor [8]

**J.** Acknowledging the physical non-interacting properties of light with the transfer medium photon in not address to issue of power loss, energy remission that there electronic counterpart had to face[3][8]

So as there's no space for any limit of barrier set to photon by nature hence there exist any one barrier that limits photons that how far a human mind can think or how far would our progress go. No matter to what extent we go photonics still photonic domain would be beyond our imagination would be allowing us to leap ahead rather far ahead. [3][5][8]

III. GENERALIZED WORKING OF SILICON BASED PHOTONIC COMPUTER

The most basic component of photonics computer is the integrated optic circuit whose implementation via different organization is all what computing is based upon considering the working of optic transistor it is considered to be compromised of six building blocks or components:- [3]. Fig 4 shows step that shall be undergone in order to utilize photon for computing.

1. A way in or a trace for light to enter.
2. Module that would be guiding the light
3. A module that would be encoding data into optical bits.
4. Interpreter of optic bits.
5. Basic computation operational components.
6. And the machine intelligence to interpret the data contained in photons

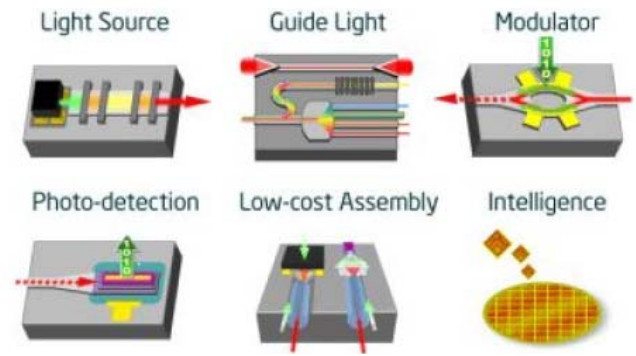

Figure 4. Modules of photonic computers. [3]

**A. Wave guides**.

Waveguides are considered to be primary concept of photonics based on principal of total internal reflection the wave guide would be related to the key properties of the matter that shall be chosen. So if we talk about implementing photonic logic on silicon we must undermine the below mentioned properties of silicon [3]. As the wave guide would be dependent on wavelength; gaped silicon which determine the shortest wavelength that would propagate through the silicon. By observations based on experimentation it was concluded that wavelength should lie between (1.3 µm – 1.6 µm). [3]

**B. Modulation**

Silicon photonic modulation is based on Mach-zester interconnectors with combination of Kerr effect and branz affect, the combination of Littoral-Maritime was latter the reflective index and when the split arrays shall recombined they either indirect constructively or destructively that we call modulation [3][7]. Fig 5 shows the modulation of laser light which is initially split, then under goes desired phase shifting and encoded into data.

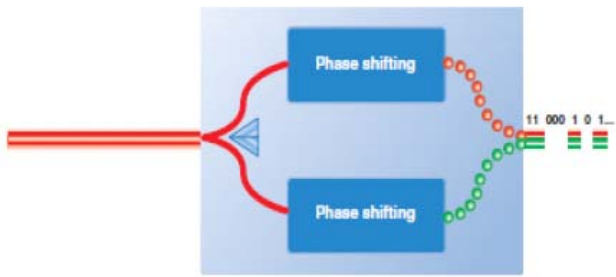

Figure 5. Modulation of light. [7]

**C. Silicon Laser**

This was a most challenging hurdle to overcome for photonics Silicon due to silicon indirect band gap, But this issue become resolve it applying principles of Roman effect and eventually a silicon laser was created based on Fabry – Perot resonator. [3]

**D. Photon Detection**

The process involves conversion of encoded data of photons to be decoded. For doing so silicon and germanium were merged. From where once data in form of photons received is forwarded to logical sector. [3]

**E. Intelligence**

Silicon has a great edge in order to be chosen under the intelligence module because of compatibility in CMOS structure. The signal received by machine intelligence and then the detector signals is interpreter. [3]

**F. Low Cost Assembly**

The components that integrate up to form a photonic chip are inexpensive but the perfection in architecture is key issue, which is ensured by facet preparation technique, precise alignment antireflective thin film location and tapes that ensure both speed and reflection. [3]

IV. APPLICATION AREAS

Photonics applications are in every aspect of our life of this modernized world. In every field of life we observe a vital contribution of photonic to make our life more comfort and reliable in our everyday used products, this was made possible by attributes of photons that yield efficiency, reliability and perfection at low cost, high speed and safety. [9]

In our every day life, we interact with each other by using telecommunication techniques, we interact with hospitals their laser surgery endoscopy and a lot for, in military applications infrared censors, command & control systems, navigation, search & rescue range and lot more in the fields of metrology, spectrometry all these things are blessing of photonics field. [9]

In short photonics could be regarded as booster for information and communication, data storage, Industrial manufacturing and ability, life science and lead it lightening and displays, security, metrology and sense, security and safety, transport, space, aeronautics rather every field of life have been boosted exponentially just due to photonics and those fields indexes its absent photonics would be soon redefining them as well, so that they could become more reliable .provides higher productivity and efficiency so that they could provide the need of up coming times. [9]

Photonics is no longer an emerging field but a present reality with multiple products readily on the market. The progress in the field of communication was possible because of its optical interconnects other fields including optical data recording, optical fibers xerography based laser printing all such wonders of present time are the practical photonics. [6][9]

Photonics benefits also motivated NASA to replace electric components by photonic counter parts in space crafts due to their greater performance, high reliability and all the above quick delivery. Though same issues are face but are expected to be resolved as photonic research is progress. [11]

Intel cooperation is also working full fledged on photonics, they have worked on silicon photonics and as a result have produced the world fastest opto-integrated chip having a speed of encoding at the rate of 200 GB/sec. [10]

Infinera cooperation has also introduced new wavelength division multiplexing wonder by photonics by making a 100 GB/Sec WDM system which is first of its type in industry.

It's a wonder which was achieved after integration over 50 discrete optics devices capable of performing distinguishes operation. Infinera is also intended to launch 100GB Ethernet on the near future. [6]

If issue of application of photonics is there it everywhere where computers are or every place where computation is taking place in short our entire life. [9]

## V. RESEARCH DIMENSIONS WITH PHOTONIC COMPUTING TECHNOLOGY

The upcoming era could be considered as an age of photonics and photonics would be a crucial element of the wonders, we would be interacting with. If we try to assume what would be our future broadband based activities the only way we see them functional is by photonic induction in them. [11][12]

The field of health sciences would be based on the photonic based operations in order to enhance rather redefines diagnosis, monitoring and treatment standards. It is expected that photons will be utilized in order to tackle the near future energy crises that would caused due to end of fossils reservoirs especially as an alternate method for electricity. If we talk about future of computation where there would be automation and artificial intelligence what would be done with the huge bulks of data that would be required to be managed and interpreted at large and at once so that we can make use of automation and artificial intelligence so that our lives are facilitated. For sure it is not sluggish electron domain then who's the answer is photon. [11][12]

Analyzing the present development and the innovations that are emerging out in the photonics domain with combination of what is being proposed there are following prediction made that are expected in photonics [9].

**A.** Enhancement in nano-scale semiconductors organic material, sub-wavelength starched as a whole referred as photonic crystals technology that would lead technologies to endless possibilities [9]. Figure 6 shows an optically operated chip

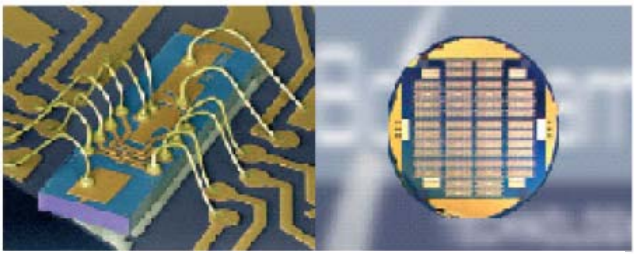

Figure 6. Laser fabricated optical chip consisting of multi grating switch capable of digitally selecting the emitted wave

**B.** Like cryptography, the other photonic phenomenon such as coherent light material integration, teleportation with the merge of other presently not known properties will yield new dimensions to photonic industry [9].

**C.** With the enhancement in nano-scale technologies and the upcoming technologies till not discovered we shall be achieving products so considered novel and for sure will be playing a vital role in widening the scope of photonics. [9]

**D.** With the further progress and enhancement in presently developed photonic products especially in the field of "silicon photonics" (CMOS Technologies) will be redefining the present domain of photonics and opening new horizons for research. [9]

**E.** It is expected that the nano optical (metal and dielectric) will reflecting their worth in the domain of memory and data, which is predicted that the breakthrough would be done by Plasmon based devices along with metal optics. All this would be done by swapping electrons with photons. [9]

**F.** All the technological products let it be a semiconductor or a loser being utilized for display purpose all are expected to be in very near future based on opto-electrical combination that would be first step towards photonic rule take over on electronic industry. [9]

Initiated by burning or heating materials, then, till the incandescent light bulb a fruit of l9th century along with combination of grid electricity boost the display and light concept to an extent no one never ever thought. Then came the era of gas discharge lamps in the middle of 20$^{th}$ century were considered as the first step towards photonics innovations in lightning and displays. [9]. Figure 7 shows the transformation of light and the future of light utilization for lightening and display purposes'.

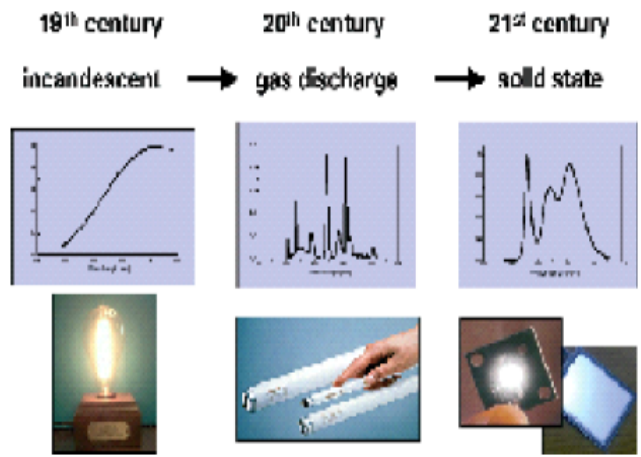

Figure 7. The transformation of light from its mode in initial stages and the future of lighting for display purposes. [9]

Now the approaching era is of solid state lightning offering a direct conversion of electricity into visible light, indeed it opened new pathways for the light solutions by proving their worth in medical diagnostics. It is expected that these solid state lightning methodology along with organic and inorganic LED's will be what the lightening future is. [9]

When we talk about display, its journey started by the invention of cathode ray tube televisions (CRT) that undergone numerous transitions and eventually we had flat panel display (FPD). [9]

The display accessories have an immense financial load over market and when we talk about future, it's going to be a lot more, that would be large, flexible but low-cost and the only we its possible is by solid state display, moreover in future time the light would be just for display, it would be performing sensing, and detection tasks or in short one can say that it has to play role in ambient intelligence as well. [9]

Photonic contribution in the field of health and life science had contributed extensively in the progress done in the field. Common example includes the breach into the world of cells and bacteria, which evolved the medical science. Let it be field of neurosurgery, ENT, oncology etc the progress in each of them is fruit of photonics. Moreover, image guided systems, laser diagnosis have also contributed their maximum to medical field evolution.

Similarly the other aspects such as endoscopes', MRI, CT scan, florescent methods all that are considered to be the pillars of modern medicine are based on photonic principles. [9] Figure 8 shows result of cell study conducted by using light. But the question arises how a photon and what is in photon that it's redefining the limit of medical sciences, the answer is as under:-

(1) Photons interact with biomaterials in real time without interacting with them. Let it be protein reactions in living cells. Even a single molecules behavior could be studied without interfering or biasing their behavior. [9]

(2) Detailed study of bio-material whether cell tissue is possible. [9]

(3) In certain field as in eye surgery or tumor removing, light could be used in form of laser and can be used for treatment purpose. [9]

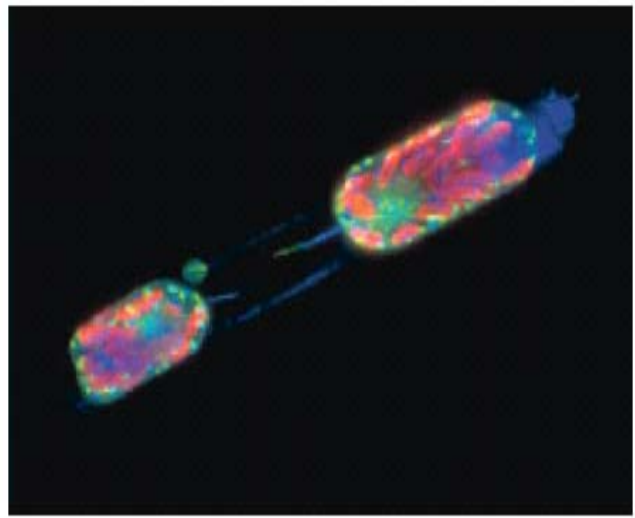

Figure 8. A result of examination of photosynthetic cell conducted by light different colors represents different cell and tissues. [9]

The mentioned characteristics made revolution by the pin-pointed identification of symptom and then the diagnostics and treatment to be followed due to exact identification of diseases. The patient specific drug would be developed, by the help of photonic based screening methods [9]. It is expected that in the near future, 3-D imaging would be used to enhance the understanding of cells, tissues there structure and presser. The diseases would be diagnosed before their apparent symptom would appear by examining the breath, saliva, urine and perspiration analysis by photonics based tools. The environment detectors would be formed the optical analysis of upcoming infection and would act as early warn systems [9]. Presently numerous technologies have been developed that utilizes photon under different conditions for numerous function. The need of present technology demands precision and accuracy without even a bit space of error. Optical sensors address this issue by guiding the process accurately and in a quicker manner with properties of each to use high speed, accuracy, feasibility along with quality to integrate with high tech automated inspection system. All this could be achieved by implement of photonic principles of 1-D and 3-D metrology, laser spectroscopy, X-ray imaging, color and pressure measurement and a lot more similar application [9]. The present of technology is intensively based on data storage which is expected to boost astronishly in the upcoming time. The scenario will lead to demand of new mechanism of data storage that holds immense huge bulks of data and for sure to build it we have no alternate platform than photonics [9]. In the field of security there are numerous ways in which photonics wonders could be utilized light, compact and efficient monitoring of environment risk, border security, traffic and weather conditions all could be by modulating different components of photonics. For security purpose remote detection of explosives, small arms, radioactive agents, drugs all could be detected from a distance. All this would be achieved by integration of optical component in security devices [9].

In the transport domain the main role is expected to be played by optical sensors. With the increasing electrical and optical components it is expected that within a certain time frame the control of transport would be handed over to optic based devices, which shall be interpreting the received data, process it and then control the behavior of object are built for e.g. the detection of pedestrian and the distance between cars travelling would be maintained by vehicles by utilizing laser sensor upcoming future would bring revolution is vehicle manufacturing industry stress sensor module, chemical sensor, occupancy sensors etc. Figure 9 depicts the laser sensing done by vehicles to detect presence of pedestrian or vehicle on road. Optics would be proving its worth by detection of pave conditions, structural sensors; all this would be redefining the transporting mode both in terms of efficiency and security [9]. The everyday used cameras, web cam etc all are based on photonics principles. Initiated as a extra ordinary technology utilized by minority it became an integral part of almost all accessories i.e. how in mobiles and laptops camera is not an option but a standard and feature, similarly, scanner previously precise in amount have not entered all hopes. In short the photo-detection has not become a vital module of present times [9].

Coming to the future, the demand is the enhanced performance as of today, with enhance form of detectors, reduce noise in the area of scanners further enhancement would be asked. When the programming style would turn a linguistic will be able to compromise on quality and the delay, definitely not then, fore sure it all not possible by electronic based architecture technological devices and then the only option left is photon. [9]

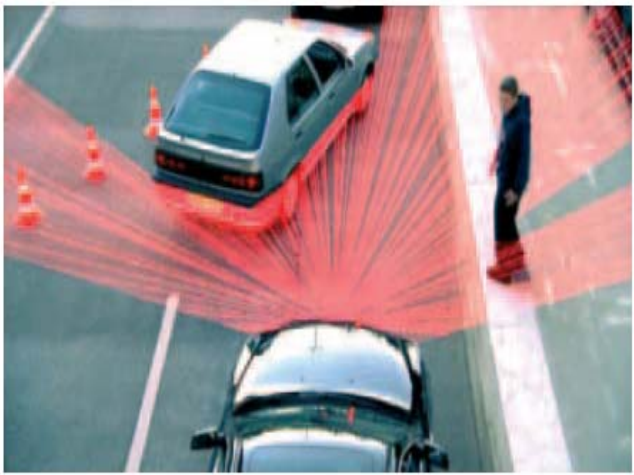

Figure 9. Sense of pedestrian and vehicles on road being done by a laser emitter and sensor operational in car.[9]

## VI. CONCLUSION

We have achieved a state that elimination of computation is not possible and nor any compromise with it is possible. The need of our future is tackling huge amount of and the only way is to have faster computing parallel processing which is only possible by introduction of photonics. Photonics is not an emerging field but a field with readily available products in market. Though some hurdles yet to over come but would be hopefully resolved and as far as its domain or future is considered questions arises how long would it last. The answer is its not a issue what amount of water on ocean holds its all about how much can you drag out of it.

### REFERENCES

[1] From Moore's Law to Intel Innovation__ Prediction to Reality; http://www.diit.unict.it/users/mpalesi/COURSES/CE_07-08/DOWNLOAD/moores-law-0405.pdf
[2] Intel turns to photonics to extent Moore's law; http://optics.org/cws/article/industry/40732
[3] Silicon Photonics; http://web.pdx.edu/~larosaa/Applied_Optics_464-564/Projects_Presented/Projects-2006/Michael_Bynum_Report_Silicon-Photonics.pdf
[4] Introducing Intel's Advances in Silicon Photonics; http://techresearch.intel.com/UserFiles/en-us/photonics/Intel_Advances_Silicon_Photonics.pdf
[5] Photonic Computing; http://nolamers.com /docs/ photonic-computing.pdf
[6] Photonic Integrated Circuits A technology and Application Primer; http://www.infinera.com/pdfs/whitepapers/Photonic_Integrated_Circuits.pdf
[7] Intel Unveils Silicon Photonics Breakthrough : High Speed Silicon Modulation; http://techresearch.intel.com/UserFiles/en-us/photonics/si02041.pdf
[8] Why Compute With Light?; http://rmrc.org/?page_id=54
[9] Towards a Bright Future for Europe; http://www.photonics21.org/download/sra_april.pdf
[10] A Record-Breaking Optical Chip; http://www.hpcwire.com/news/A_Record-Breaking_Optical_Chip.html
[11] Recent Photonic Activities Under the NASA Electronic parts and Packaging(NEPP) Program; http://trs-new.jpl.nasa.gov/dspace/bitstream/2014/9429/1/02-1649.pdf
[12] The European Technology Platform for Photonics; http://www.photonics21.org/download/Photonics21FlyerGeneralInformation.pdf

**AUTHORS PROFILES**

**M. Aqeel Iqbal**

M. Aqeel Iqbal Is An Assistant Professor In The Department Of Software Engineering, Faculty Of Engineering And Information Technology, Foundation University, Institute Of Engineering And Management Sciences, Rawalpindi, Pakistan. As A Researcher He Has A Deep Affiliation With The College of E & ME, National University Of Sciences And Technology (NUST), Rawalpindi, Pakistan.

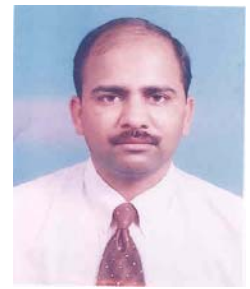

**Umer Farooq, Muhammad Ahsan and Mashood Malik**
    Umer Farooq, Muhammad Ahsan and Mashood Malik Are Students Of BCSE Program Of The Department Of Software Engineering, Faculty Of Engineering And Information Technology, Foundation University, Institute Of Engineering And Management Sciences, Rawalpindi, Pakistan.